\documentclass[a4,aps,showpacs,preprintnumbers,amsmath,amssymb]{revtex4} 
\usepackage{dcolumn}
\usepackage{graphicx}
\usepackage{amsfonts} 
\usepackage{epstopdf}
\usepackage{color}
\usepackage{here}
\usepackage{verbatim}

\begin{document} 
\title{Generation and decay of persistent current in a toroidal Bose-Einstein condensate}
\date{\today}
\author{A.I. Yakimenko$^{1}$, S.I. Vilchinskii$^{1,2}$,  Y. M. Bidasyuk$^{3,4}$, Y.I. Kuriatnikov$^1$,  K.O. Isaieva$^1$, M. Weyrauch$^{3}$}
\affiliation{$^1$Department of Physics, Taras Shevchenko National University of Kyiv, Volodymyrska Str. 64/13, U-01601, Kyiv, Ukraine \\
$^2$Departement de Physique Theorique Center for Astroparticle Physics,
Universite de Geneve, Quai E. Ansermet 24, 1211 Geneve 4, Switzerland \\
$^3$Physikalisch-Technische Bundesanstalt, Bundesallee 100, D-38116 Braunschweig, Germany \\
$^4$Bogoliubov Institute for Theoretical Physics, Metrolohichna str. 14b, U-03680, Kyiv, Ukraine}
\begin{abstract}
Persistent current, or ``flow without friction'', as well as quantum vortices are the hallmarks
of superfluidity. Recently a very long-lived persistent flow of atoms has been experimentally
observed in Bose-Einstein condensates trapped in a
ring-shaped potential. This enables fundamental studies of
superfluidity and may lead to applications in high-precision
metrology and atomtronics. We overview our recent theoretical studies of the generation and decay of the persistent current in a toroidal atomic Bose-Einstein condensate, and discuss our new investigation of the hysteresis in the atomtronic circuit.
\end{abstract}
 \maketitle

\section{Introduction}
The dramatic progress in the field of Bose--Einstein condensates (BECs) of atomic gases continuing last two decades is driven by a combination of new experimental techniques and theoretical
advances. Nowadays BECs have become an ultralow-temperature laboratory for nonlinear physics, many-body physics and condensed matter physics, exhibiting superfluidity, quantized vortices, solitons, Josephson junctions and quantum phase transitions.
The achievement of Bose--Einstein condensation in dilute atomic gases provided the
opportunity to observe and study superfluidity in an extremely clean and well-controlled
environment and to search for the subtle links between superfluidity and Bose-Einstein condensation.

Superfluids are distinguished from ordinary fluids by their ability to support
dissipationless flow, which is often called {\it persistent current.}
The existence of a persistent current is related to a stable quantized
vortex, which is a localized phase singularity with integer winding numbers. Generation and decay of a persistent current is governed by dynamics of these quantum vortices.

Recently, persistent flow of atoms has been observed in BECs trapped in a ring-shaped potential
 \cite{PhysRevLett.99.260401,RamanathanPhD,PhysRevLett.106.130401,PhysRevA.86.013629,
Beattie13, PhysRevLett.110.025302,Wright2013,PhysRevLett.111.205301,
 PhysRevLett.111.235301,PhysRevLett.113.045305,PhysRevLett.113.135302,
 Ryu2014NJPh,nature14}. With a system of laser beams it is
possible to create a toroidal condensate
and a repulsive barrier rotating around the
ring. The presence of a barrier produces a localized region
of reduced superfluid density in the condensate  annulus -- a weak link.
A super-fluid ring with a weak link 
as well as a super-conducting circuit with a Josephson junction
can act as a nonlinear interferometer,
allowing the construction of high-precision detectors
such as superconducting quantum interference
devices (SQUID), magnetometers and superfluid gyroscopes.
The fast response of the condensate phase winding on the variations in the angular velocity of the barrier is analogous to
that in the superconducting current of the SQUID caused by changes in the external
magnetic field.  This feature makes the atom SQUID ideal for future highly sensitive
rotation-measurement devices \cite{PhysRevLett.111.205301}. 

Ring-shaped Bose-Einstein condensates in toroidal traps are the subject of many experimental and theoretical investigations
\cite{PhysRevA.66.053606,BenakliEuL99,Brand01,Das2012, PhysRevA.64.063602,PhysRevA.74.061601}
which study persistent currents \cite{PhysRevLett.99.260401,Beattie13,PRA2013R}, weak links \cite{PhysRevLett.106.130401,PhysRevLett.110.025302}, formation of matter-wave patterns by rotating potentials
\cite{PhysRevA.86.023832, LiMalomed2},
solitary waves \cite{Brand01,Berloff09}, and the decay of the persistent current via abrupt change of the rotation state (phase slips) \cite{PhysRevA.86.013629, PhysRevA.80.021601,Piazza2013}.
A persistent atomic flow in a toroidal trap can be created  by transferring  angular momentum from optical fields \cite{PhysRevLett.99.260401,PhysRevLett.110.025302} or by stirring with a rotating barrier~\cite{PhysRevLett.110.025302,Wright2013}.
 These experiments has shown that toroidal geometry makes
it possible to avoid the fast vortex splitting that takes place in
a singly connected BEC and study the properties of vortices
with large winding number.

  The present paper is organized as follows.
 In Section \ref{spinor} we briefly review a theoretical treatment \cite{PRA2013R} of the remarkable effect in a spinor toroidal BEC, which was recently discovered experimentally in Cavendish laboratory \cite{Beattie13}. In Section \ref{smallbeam} we discuss the analysis of the vortex excitations in a quasi-two-dimensional annular BEC reported in \cite{SmallWL_arxiv14}. This work was inspired by the experimental demonstration \cite{Wright2013} of vortices in a toroidal trap, which are excited using a  \emph{small} (diameter less than the width of the annulus) variable-height potential barrier (a ``stirrer''). In Section \ref{sec_weaklink} we briefly overview our recent work \cite{yakimenko2014vortices}, which was motivated by another experiment  \cite{PhysRevLett.110.025302}, where the deterministic phase slips driven by a \emph{wide} stirrer (rotating weak link) were demonstrated.
  In Section \ref{sec_hysteresis} we present our new theoretical results concerning hysteresis effect in two-dimensional ring-shaped condensate.
  In Section \ref{Conclusions} we make general conclusions.

\section{Stability of persistent currents in spinor Bose-Einstein condensates}\label{spinor}
Spinor Bose-Einstein condensates are multicomponent BECs with additional spin degree of freedom. Due to nonsymmetric nature of intercomponent interaction they can demonstrate particularly rich variety of phenomena (for the recent review see \cite{UedaPhysRep12,RevModPhys.85.1191}).
  For two-component BECs in 1D or 2D traps, the stability of the persistent currents and their decay mechanisms  were under investigation in Refs. \cite{SmyrnakisPRL09,SmyrnakisPRA10,SmyrnakisPRA10_annular,SmyrnakisPRA07,JacksonKavoulakisPRA06,KavoulakisJLTP07,BrtkaMalomedPRA10}.
Smyrnakis {\it et al.}~\cite{SmyrnakisPRL09} concluded that in a strictly one-dimensional ring persistent currents with  circulation larger than one  are stable only in single-component gases. However, recently ~\cite{ANO13}, this conclusion was challenged, and it was claimed that persistent currents at higher angular momenta may be stable in certain parameter regions of a two-component system in a 1D ring.
In Ref.  \cite{BargiKavoulakis10} it was found on the basis of mean field calculations and supported by exact diagonalization results, that  persistent currents in 2D traps may be stable under specific conditions.

 Experimentally such system was investigated recently~\cite{Beattie13} for a toroidally trapped gas of $^{87}$Rb atoms in two different spin states.
It was discovered in this  experiment that  the supercurrent is unstable if the spin polarization is below a well-defined
critical value and was shown  that only the magnitude of the spin-polarization vector, rather than its orientation in spin space, is relevant for supercurrent stability. In Ref. \cite{PRA2013R} we have suggest a theoretical explanation of this remarkable effect.

The main properties of an ultracold dilute spin-1 atomic BEC at zero temperature can be accurately described by a set of mean-field Gross-Pitaevskii  equations (GPEs) for the spin components $(\Psi_+,\Psi_0,\Psi_-)$ of the three-component order parameter $\vec\Psi$. These equations correspond to the Hamiltonian $H=H_0+H_A$ with the spin-independent part (see, e.g. ~\cite{Matuszewski12})
\begin{equation}\label{hamilt1}
H_0=\sum_j\int \Psi_j^*\left(-\frac{\hbar^2}{2M}\Delta+\frac{c_0}{2}n+V(\textbf{r})\right)\Psi_j  d^3\textbf{r},
\end{equation}
where $\textbf{r}=(x,y,z)$ and $n=n_++n_0+n_-=\sum_j|\Psi_j|^2$ is the total density; the strength of interaction between atoms $c_0=\frac{4 }{3}\hbar^2\pi(2a_2+a_0)/M$ is given in terms of the s-wave scattering length $a_S$ for atom pairs with total spin $S$.
The spin-dependent part $H_A$ of the Hamiltonian is given by
\begin{equation}\label{hamilt2}
H_A=\int \left(E_+n_++E_-n_-+E_0n_0+\frac{c_2}{2}|\textbf{F}|^2\right) d^3\textbf{r}
\end{equation}
where $E_j$ is the Zeeman energy of the state $\Psi_j$, $c_2=\frac{4 }{3}\hbar^2\pi(a_2-a_0)/M$, and $\textbf{F}$ is the spin density:
$$\textbf{F}=(F_x, F_y, F_z)=(\vec\Psi^\dag \hat F_x \vec\Psi, \vec\Psi^\dag \hat F_y \vec\Psi, \vec\Psi^\dag \hat F_z \vec\Psi),$$
where $\hat F_{i}$ are the spin-1 matrices and $\vec \Psi=(\Psi_+, \Psi_0, \Psi_-)$.
\begin{figure}[ht]
  \includegraphics[width=0.6\textwidth]{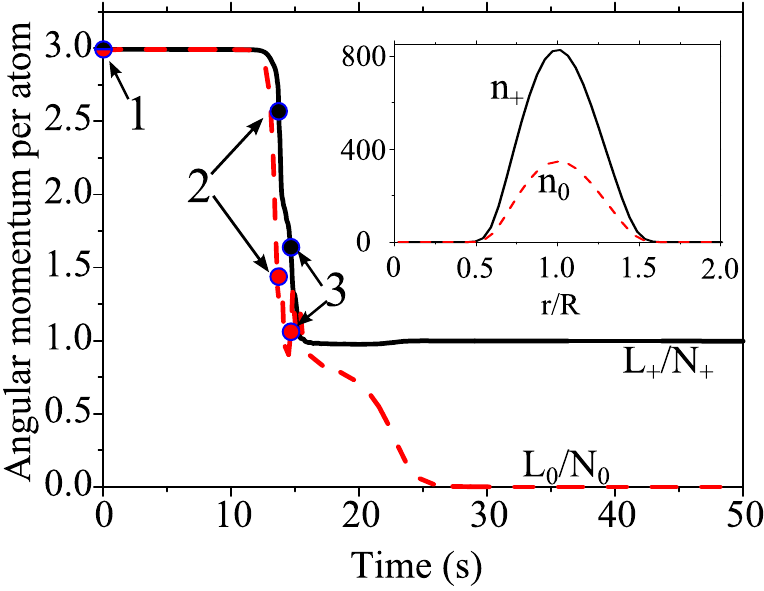}
  \caption{(Color online) Decay of the persistent current in a two-component condensate with spin-population imbalance $P_z=(N_+-N_0)/(N_++N_0)=0.4$. Shown are the angular momentum per atom for spin component $m_F=+1$ (solid black curve) and $m_F=0$ (dashed red curve).  The inset represents the initial radial distributions of the 2D densities $|\psi_+|^2$ and $|\psi_0|^2$ for the triply charged persistent current in the toroidal trap. The integer numbers indicate the moment of time for the snapshots shown in Fig. \ref{phase} (b). \emph{Source}: From paper by author \cite{PRA2013R}.}
  \label{angularMomentum}
\end{figure}
\begin{figure}[ht]
  \includegraphics[width=0.6\textwidth]{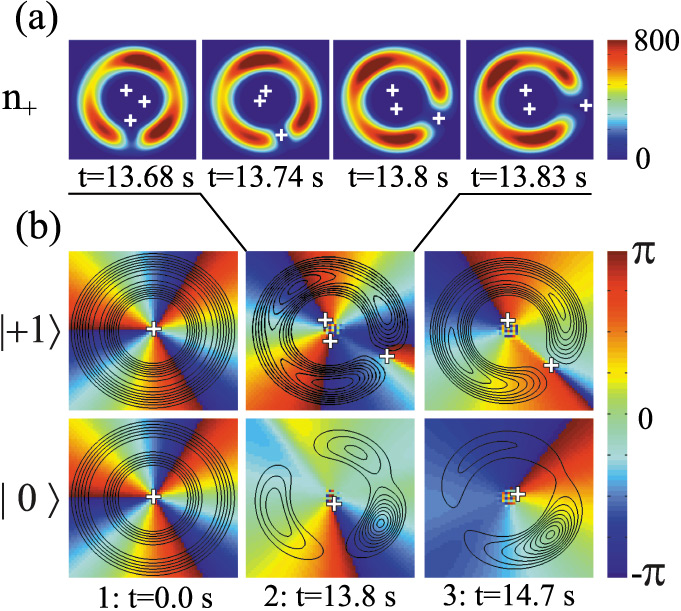}
  \caption{(Color online) (a) The detailed dynamics of a phase slip from charge-$3$ vortex to a charge-2 vortex. The color code shows the density of the $m_F=+1$ spin component. The positions of the vortex cores are indicated by crosses. (b) Color-coded phase of two spin components combined with density isolines during phase-slips at the times indicated in Fig. \ref{angularMomentum} by the integers.  \emph{Source}: From paper by author \cite{PRA2013R}.}
  \label{phase}
\end{figure}

In accordance with the experimental set-up of Ref.~\cite{Beattie13}, we approximate the external toroidal optical trapping potential $V(\textbf{r})$ by a superposition of a harmonic potential with trapping frequency $\omega_z$ (which models the elliptic highly anisotropic ``sheet'' beam) and a radial Laguerre-Gauss potential (which models the ``tube'' beam):
\begin{equation}
  V(\textbf{r})=\frac{M\omega_z^2 z^2}2-V_0\left(\frac{r}{R}\right)^{2q} e^{-q\left(r^2/R^2-1\right)},
\end{equation}
where $M$ is the atomic mass, $R$ is the radius of the trap (radial coordinate of the minimum of the potential), $r=\sqrt{x^2+y^2}$, and $V_0$ is the trap depth. 
The parameter $q=3$ corresponds to the topological charge of the optical vortex which produces the Laguerre-Gauss potential.

The harmonic potential creates a tight binding potential in $z$ direction, so that the BEC cloud is ``disk-shaped'' ($l_z\ll R$ with $l_z=\sqrt{\hbar/(M\omega_{z})}$ the longitudinal oscillator length). Consequently, we assume that the longitudinal motion of condensate is frozen in:
\begin{equation}\label{factorization}
\Psi_j(\mathbf{r},t)=\tilde\Psi_j(r,t)\Upsilon(z,t),
\end{equation}
 where
$\Upsilon(z,t)=(l_{z}\sqrt{\pi})^{-1/2}\exp(-\frac{i}{2}\omega_zt-\frac12z^2/l_{z}^2)$. After integrating out the longitudinal coordinates in the  GPEs corresponding to the Hamiltonian $H$ given in Eq.~(\ref{hamilt1}) and (\ref{hamilt2}), and accounting for dissipative effects (see below), we obtain a  set of three coupled differential equations in 2D:
\begin{eqnarray}
  (i-\gamma)\frac{\partial \psi_\pm}{\partial
  t}&=&\hat{\mathcal{H}}_\pm\psi_\pm+\nu_a \psi_0^2\psi^*_\mp,\label{main1}\\
  (i-\gamma)\frac{\partial \psi_0}{\partial
  t}&=&\hat{\mathcal{H}}_0\psi_0+2\nu_a\psi_+\psi_-\psi^*_0,\label{main2}
\end{eqnarray}
with
\begin{eqnarray}
\hat{\mathcal{H}}_\pm&=&-\frac{1}{2}\Delta_\perp-\mu_\pm+V(r)+\nu_s n +\nu_a(n_0+n_\pm-n_\mp), \nonumber\\ \hat{\mathcal{H}}_0&=&-\frac{1}{2}\Delta_\perp-\mu_0+V(r) -\epsilon +\nu_s n +\nu_a (n_++n_-).\nonumber
\end{eqnarray}
Here we have introduced  the 2D Laplace operator $\Delta_\perp$ and dimensionless space and time coordinates $r\rightarrow r/R$, $z\rightarrow z/l_z$, $t\rightarrow \Omega_0 t$ with $\Omega_0=\hbar/({M R^2})$ as well as the dimensionless chemical potential $\mu_j\to \mu_j/(\hbar\omega_r)$ of the  spin component $j$.   The order parameter takes the form $\psi_j e^{-i\mu_j t}=\tilde\Psi_j/C$ with $C^2=\sqrt{2\pi}\hbar\omega_r l_z/c_0$.
Furthermore we define $V(r)=-V_0 r^{2q}\exp{(-q(r^2-1))}$,  $\nu_s=\textrm{sgn} (c_0)=+1$, and  $\nu_a=c_2/c_0=-4.66\cdot 10^{-3}$ for a $^{87}$Rb condensate.

\begin{figure}[ht]
 \includegraphics[width=0.6\textwidth]{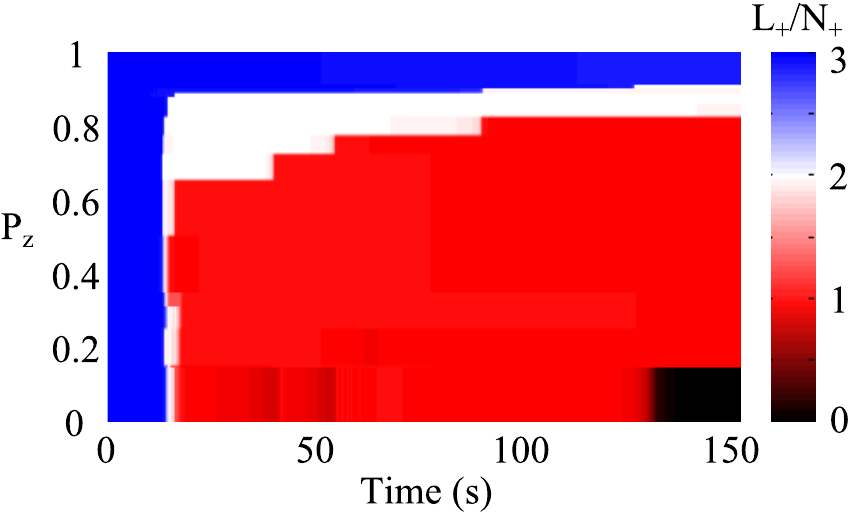}
  \caption{(Color online) Angular momentum per atom $L_+/N_+$ as a function of time and $P_z$. Angular momentum is represented by the color code and characterizes vortex charge if $L_+/N_+$ is integer. It can be seen that there is a very sharp edge between stable and unstable currents at about $P_z\approx0.89$.  \emph{Source}: From paper by author \cite{PRA2013R}.}
  \label{Pz}
\end{figure}

Non-equilibrium dissipative effects are of crucial importance since they provide the mechanism for the damping of the vortices.
We describe such effects by the phenomenological parameter $\gamma>0$ on the left hand side of GPEs (\ref{main1}) and (\ref{main2}).
Dissipation means that energy and particles are exchanged between the condensate and a thermal reservoir. Here, we use the phenomenological approach introduced by Choi et al. \cite{Choi98} for atomic BEC in a manner similar to that originally proposed by Pitaevskii \cite{Pitaevskii59}.

The GPEs (\ref{main1}) and (\ref{main2}) with $\gamma>0$ conserve neither the energy nor the number of particles. To simulate realistic experimental behavior of particle number we adjust the chemical potential $\mu(t)$ at each time step so that the  number of condensed particles slowly decays with time, $N(t)=N(0)e^{- t/\tau_0}$, in accordance with measurements for lifetime of atomic BECs reported in Ref.~\cite{Beattie13}.

We numerically simulate the dynamics of the persistent flow in a toroidal trap using the two-dimensional dissipative mean-field model given by Eqs.~(\ref{main1}) and (\ref{main2}). It turns out that due to the weakness of the spin-dependent part of the interaction and rather significant dissipation, the population $N_-$ remains very low and can be neglected.

During a phase slip the angular momentum is not integer, as is clearly seen from Fig. \ref{angularMomentum} which presents a typical example of the temporal dynamics of $m_F=+1$ and $m_F=0$ spin components of the order parameter.
As we observe during a phase slip, the density distribution of the two spin components is highly anisotropic
(see Fig.~\ref{phase}(a)). It turns out that the smaller the admixture of the minor component is the more time is needed for the development of the symmetry-breaking azimuthal instability, and thus the longer is the lifetime of the persistent current. This is of no surprise, since the phase separation appears as a result of a repulsive nonlinear inter-component interaction. The phase separation grows out of small azimuthal perturbations which are energetically favored in a condensate with comparable number of atoms in both components.

It is obvious from Fig. \ref{phase}(b) that as the result of the phase separation, the regions with reduced density in one component are filled by atoms of the other component (so that the total density distribution $n=n_++n_0$ remains axially-symmetric). These dynamical weak links play a key role for the decay of the persistent current: The vortex line, which is trapped inside the internal toroidal hole, can easily drift through these ``gates'', which consequently changes the topological charge by one unit.

This hypothesis is supported by investigations of the dynamics of the phase of the order parameter and by analysis of the vortex core position during the phase slip. Typical examples of phase-slips are shown in Fig. \ref{phase}, where the white crosses indicate the positions of the vortex cores.

To compare our theoretical findings with the experiment \cite{Beattie13} we present the angular momentum per particle in the major component as the function of the spin-population imbalance $P_z=(N_+-N_0)/(N_++N_0)$ and time. In agreement with the experiments it is seen from Fig. \ref{Pz} that above a well-defined (``critical'') value for $P_z$ the persistent current is stable. Below this critical value the supercurrent decays rapidly. However, the
experimentally observed ``critical'' value ($P_z\approx 0.64$) is well below our predictions ($P_z\approx 0.89$). The difference between our theoretical estimates and experimental results could be minimized, if one properly accounts for possible deviation in the experimental conditions \cite{PhysRevA.86.013629} of 3D distribution of the trapping potential from the pure Laguerre-Gauss shape assumed in our 2D model.   A more accurate description of the dissipative processes could also improve the quantitative agreement with the experiment. Indeed, the phenomenological parameter $\gamma$, which describes the dissipation in our simulations, is in general temperature- and position-dependent, which is not included in our simple model.
\section{Generation of the persistent current in a stirred toroidal BEC}
 \begin{figure}[ht]
  \includegraphics[width=0.7\textwidth]{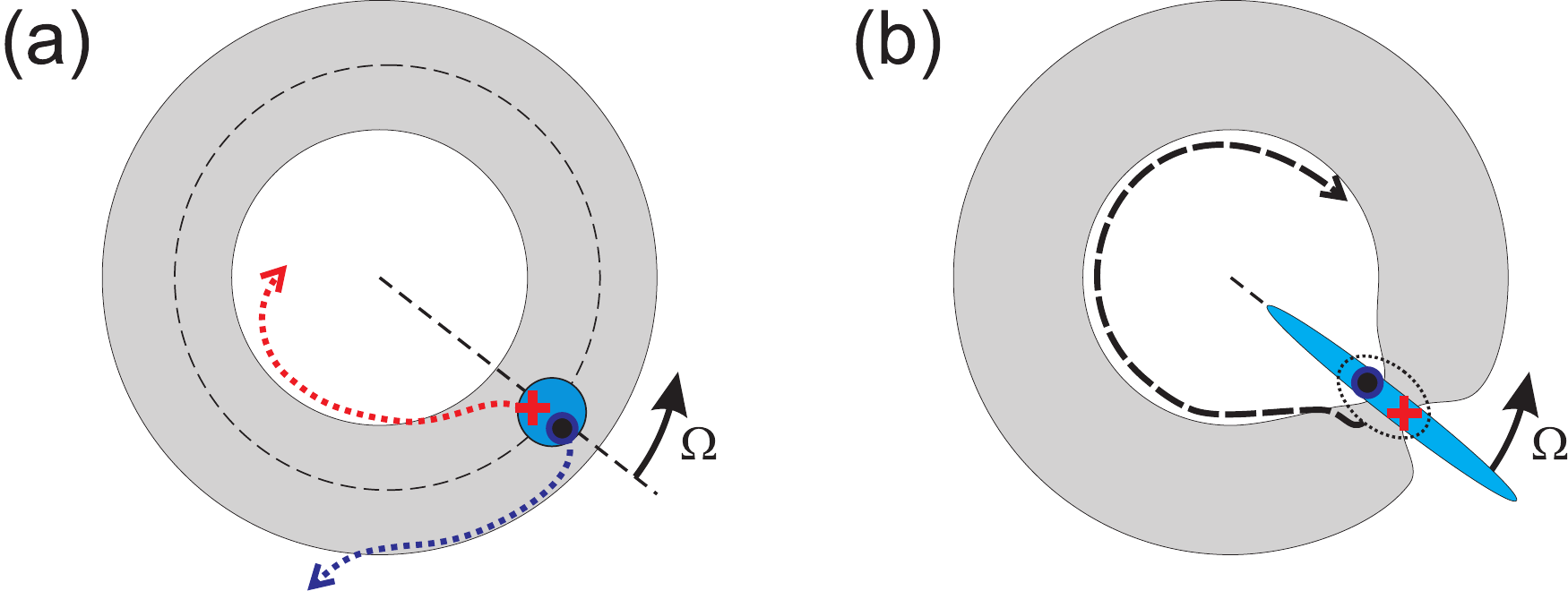}
  \caption{(Color online) Schematics of two different scenarios of the persistent current generation in toroidal BEC by rotating blue-detuned laser beams: (a) A small (diameter less than the width of the annulus) stirrer was used in the experiment \cite{Wright2013}. In this case a vortex-antivortex pair is nucleated near the center of the rotating barrier in the bulk of the condensate  \cite{SmallWL_arxiv14}. Then the pair undergoes a breakdown and the antivortex (blue circle) moves spirally to the external surface of the condensate and finally decays into elementary excitations, while the vortex (red cross) becomes pinned in the central hole of the annulus. (b) A phase slips driven by rotating weak were observed in the experiment  \cite{PhysRevLett.110.025302} and have been investigated theoretically in Ref. \cite{yakimenko2014vortices}.  The vortex line from the external periphery and the anti-vortex line from the internal region approach each other and create a vortex-antivortex dipole. This coupled pair of vortices circles clockwise inside the central hole of the toroidal condensate.
}
  \label{SchematicStirring}
  \end{figure}

\subsection{Persistent current generation by a small stirrer}\label{smallbeam}
Very recently \cite{Wright2013}, vortices in a toroidal trap were excited using a  \emph{small} (diameter less than the width of the annulus) variable-height potential barrier (a ``stirrer'') with an angular velocity ranging from zero up to the speed of sound in the condensate. A wide range of the experimental parameters used in Ref. \cite{Wright2013} opens an intriguing possibility for theoretical investigation of excitation of the persistent current in a stirred ring-shaped BEC at different regimes.
As in case of decay of persistent flow, in modeling of nucleation
of vortices dissipative effects are of crucial importance
since they provide the mechanism for damping of elementary
excitations in the process of relaxation to an equilibrium state.
It is the dissipation that either causes the vortex line to drift
to the outer edge of the condensate (where vortices decay) or leads
to the pinning of the vortex in the central hole of the ring-shaped condensate. The relaxation of the vortex core position to the
local minimum of the energy leads to formation of the metastable persistent current.

The trapping potential
$$V(\textbf{r})=V_{\textrm{tr}}(r,z)+V_{\textrm{b}}(x,y,t),$$
consists of an axially-symmetric time-independent toroidal trap:
\begin{equation}\label{PotentialToroidal}
V_{\textrm{tr}}(r,z)=\frac12 M\omega_z^2 z^2+\frac12 M\omega_r^2(r-R)^2,
\end{equation}
where $r=\sqrt{x^2+y^2}$,
and the repulsive potential of optical blue-detuned stirring beam:
\begin{equation}\label{V_beam}
V_{\textrm{b}}(x,y,t)=f(t)e^{-\frac{1}{2 d^2}\left\{[x-x_0(t)]^2+[y-y_0(t)]^2\right\}},
\end{equation}
where $\left\{x_0,y_0\right\}=\left\{R\cos(\Omega t),R\sin(\Omega t)\right\}$ is the coordinate of the barrier center, moving counter-clockwise with a constant angular velocity $\Omega$ through the maximum of condensate density [see Fig. \ref{SchematicStirring} (a)]. According to experimental parameters the effective width of the beam $d$ is  less than the width of the annulus $\Delta R$, but is ten times greater than the healing length $\xi$.  The function $f(t)$ describes the temporal  changes in the barrier height: $f(t)$ linearly ramps up  in the first $0.1$~s of evolution from zero to $U_b$ and remains unchanged for $0.8$~s, then in the last $0.1$~s it ramps down to zero again.

Similarly to how it was done for GPE for spinor condensates [see Eq. (\ref{factorization})] one can obtain a dissipative dimensionless GPE in 2D:
\begin{equation}\label{GPE_dissipative1}
(i-\gamma)\frac{\partial \psi}{\partial t} = \left[-\frac{1}{2} \Delta_\perp + V(x,y,t)+  g|\psi|^2 -\mu\right]\psi,
\end{equation}
where $V(x,y,t)=\frac12(r-R)^2+V_\textrm{b}(x,y,t)$ and  $R$ are the dimensionless potential and the radius of the trap,
$g=\sqrt{8\pi}a_s/l_z=1.54\times 10^{-2}$ is the dimensionless 2D interaction constant, $\psi\rightarrow l_r \psi$ is dimensionless wave-function.
Here we use  harmonic oscillator units: $t\to \omega_r t$, $(x,y)\to (x/l_r,y/l_r)$, $V_\textrm{b}\to V_\textrm{b}/(\hbar\omega_r)$, $\mu\to \mu/(\hbar\omega_r)$.

In Ref. \cite{SmallWL_arxiv14} we have observed two regimes of the vortex excitation in a toroidal BEC.
For a {\em low angular velocity} of the stirrer vortex-antivortex pair is nucleated near the center
of the barrier (see Fig. \ref{SchematicStirring} (a)). Then the pair undergoes a breakdown and the antivortex moves spirally to the external surface of the condensate and finally decays into elementary excitations, while the vortex becomes pinned in the central hole of the annulus adding  a unit to the topological charge of the persistent current. The progressive drift of the vortices towards the external condensate boundary is the result of dissipation, which leads to the vortex energy decay. Described behavior is characteristic for low $U_b$ intensities. In contrast, if $U_b$ is much higher than the threshold for vortex nucleation, then a wide weak link develops in a ring and breaks the potential barrier for the external vortices. As a result, a vortex from the outside of the ring can enter through the weak link. This mechanism is similar to stirring with a wide barrier observed in Ref. \cite{PhysRevLett.110.025302} and described theoretically in Ref. \cite{yakimenko2014vortices}.

For {\em higher angular velocities} of the stirrer the dominating source of vortices is the instability of the external surface modes. First, ripples appear at the external surface, and then several vortices nucleate simultaneously. It is remarkable that Bogoliubov analysis gives practically the same $\Omega_c$ above which such surface modes are exited. Also, similar to the case of high barrier intensities $U_b$, vortex lines come into the bulk of the condensate through the rotating weak link. Further complex dynamics of the vortices is governed not only by the condensate inhomogeneity and dissipation effects, but also by the interplay between condensate flows corresponding to other vortices. The number of vortices increases dramatically in our simulations with increasing barrier amplitude $U_b$, so that dynamics of the vortices becomes quite irregular.

As is well known (see, e.g. \cite{Tsubota13}), small-scale forcing may generate large-scale flows in effectively 2D turbulent classical and quantum fluids. The energy of small-scale forcing transits first to irregular vortex distributions and then relaxes to a large-scale flow in a form of a circulating superflow. For example, as is shown in Ref. \cite{SmallWL_arxiv14}, the number of vortices and antivortices rapidly increases when the   stirring beam ramps up, and decays to zero when the stirring beam is switched off, while  the angular momentum per atom saturates to an integer number $m$ corresponding to the $m$-charged persistent current.

It turns out that the results of Ref. \cite{SmallWL_arxiv14} are in a good agreement with the experimentally measured \cite{Wright2013} threshold for vortex excitation especially for $\Omega>\Omega_c\approx 16$ Hz. For the lower angular velocities the experimentally measured threshold value of the barrier height $U_b$ is well below the predictions of our 2D model, as well as of the 1D model suggested in Ref. \cite{Wright2013}.

\subsection{Quantum vortices driven by a rotating weak link}\label{sec_weaklink}
In the present subsection we briefly overview our recent investigation \cite{yakimenko2014vortices} of the formation of  persistent currents by a wide rotating barrier, i.e. the barrier that is wider than the ring [see Fig. \ref{SchematicStirring} (b)].
The rotating barrier dynamically modifies the condensate density. Then it not only creates a weak link, but also induces superflows in the ring-shaped condensate.
We have analyzed in Ref. \cite{yakimenko2014vortices} the influence of both these factors on the energetic and dynamical stability of the vortices in the toroidal condensate.

In the mean field approximation, the dynamics of a system of weakly interacting degenerate atoms close to
thermodynamic equilibrium and subject to weak dissipation is described by the following dimensionless 3D GPE
\cite{Pitaevskii59,Choi98}:
\begin{equation}\label{GPE_dimensionless}
(i-\gamma)\frac{\partial \psi}{\partial t} = \left[-\frac{1}{2} \Delta + V(\textrm{\textbf{r}},t) +  g|\psi|^2 -\mu\right]\psi.
\end{equation}
The external potential $V(\textrm{\textbf{r}},t)=V_t(r,z)+V_b(\textrm{\textbf{r}},t)$ consists of the axially-symmetric, time-independent toroidal trap (\ref{PotentialToroidal}) and the time-dependent potential of a rotating repulsive barrier $V_b(\textrm{\textbf{r}},t)$.
The rotating barrier is experimentally created by a blue-detuned laser beam scanning across the condensate in radial direction, with a
scan amplitude greater than the width of the annulus~\cite{PhysRevLett.110.025302}.
For simplicity, we have assumed the resulting averaged potential to be homogeneous in radial direction across the toroidal condensate,
\begin{equation}\label{straigthBeam}
V_b(\textbf{r}_\perp,t)=U(t)\Theta(\textbf{r}_\perp\cdot \textbf{n})e^{-\frac{1}{2c^2}\left[\textbf{r}_\perp\times \textbf{n}\right]^2},
\end{equation}
where  $\textbf{r}_\perp=\left\{x,y\right\}$ is the radius-vector in $(x,y)$ plane and the unit vector $\textbf{n}(t)=\left\{\cos(\Omega t),\sin(\Omega t)\right\}$  points along the azimuth of the barrier maximum. The Heaviside theta function $\Theta$ in Eq.~(\ref{straigthBeam}) assures a semi-infinite radial barrier potential starting at the trap axis. All parameters of the problem were chosen to match the experimental conditions in~\cite{PhysRevLett.110.025302} (see Ref. \cite{yakimenko2014vortices} for details).

Following the experimental setup~\cite{PhysRevLett.110.025302} we first prepare a nonrotating state in the toroidal trap. We use the imaginary time-propagation method to find numerically the stationary solution of the Eq.~(\ref{GPE_dimensionless}) without the weak link ($V_{b}=0$) and without dissipation ($\gamma=0$).
In the experimental procedure the amplitude $0\le U(t)\le U_b$ of the weak link  varies in time: the height of the potential barrier ramps up to a maximum value $U_b$
during the first 0.5\,s. For the next 0.5\,s the amplitude of the barrier remains constant, and then, the potential barrier ramps down again within 0.5\,s. During all the 1.5\,s of stirring time the barrier moves anti-clockwise with constant angular velocity $\Omega$.
In the experiment \cite{PhysRevLett.110.025302} the values of angular velocity were taken in the range $\Omega / 2\pi \le 3\,$Hz, that correspond to linear velocities well below the speed of sound propagating around the ring. For slow rotation rates pronounced quantized phase slips were observed at  well-defined critical angular velocities.
In Ref. \cite{yakimenko2014vortices} we have numerically simulated the generation of persistent currents via  such phase slips driven by a wide rotating barrier using parameters adjusted to the experimental conditions in Ref.~\cite{PhysRevLett.110.025302}.
 \begin{figure}[ht]
  \includegraphics[width=0.5\textwidth]{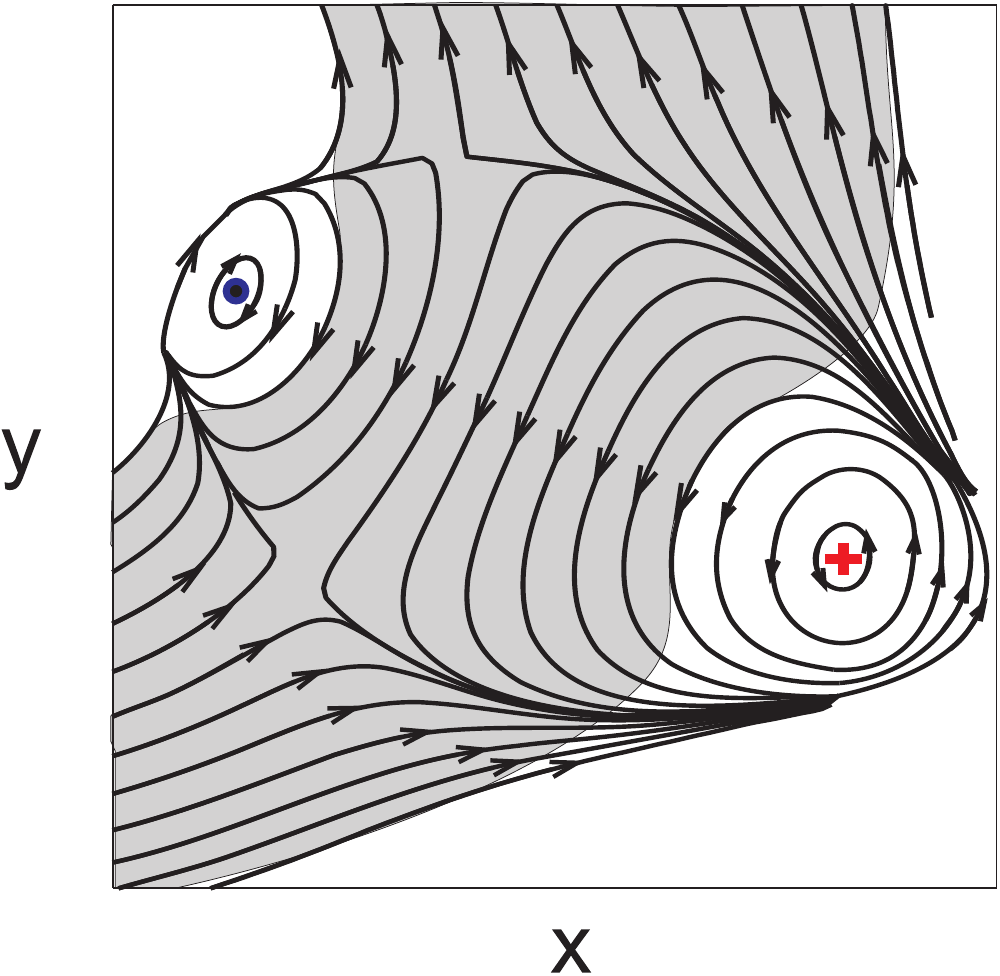}
  \caption{(Color online) Schematics of the  superflow stream lines close to the weak link, which rotates anti-clockwise. The position of the vortex (antivorex) is indicated by red cross (blue circle). }
  \label{StremLines}
  \end{figure}

Let us discuss the mechanism of these phase slips. During the stirring procedure two vortices of opposite charges are nucleated on inner and outer edges of the weak link (see Fig. \ref{StremLines}). As the superflow velocity is bigger on the outer edge, the vortex from outside enters the weak link first and travels towards the central hole as illustrated in Fig. \ref{SchematicStirring} (b).
While the vortex traverses the weak link, a negatively-charged anti-vortex line from the central hole and the incoming vortex approach each other and create a vortex-antivortex dipole. This coupled pair of vortices circles clockwise inside the central hole
until it reaches the region of the weak link.  The moving dipole usually escapes from the central hole and finally decays, however, during the stirring time the vortex dipole can  enter and leave several times  the central hole of the ring through the weak link.

If the barrier rotation rate is well above a threshold, usually several vortices enter into the central hole and form bound pairs with anti-vortices from the central hole. Finally the vortex-antivortex pairs jump out of the condensate. The total angular momentum of the escaping vortex dipole is equal to zero, but each time an external vortex enters the condensate it adds one unit of topological charge to the global persistent current.

 For higher angular velocities some integer-valued fluctuations appear in $L_z/N$ as the function of $\Omega$.  When $\Omega$ grows further the angular momentum increases on average, however the specific value of $L_z/N$ at $t=1.5$ s becomes quite unpredictable. Surprisingly, even substantial increase of the evolution time to 4.5\,s does not remove all the fluctuations. The reason is that the interaction between flows of annular vortices together with moving barrier leads to a quite complicated dynamics which can not be considered as a deterministic phase slips any more.

\begin{figure}[ht]
   \includegraphics[width=0.6\textwidth]{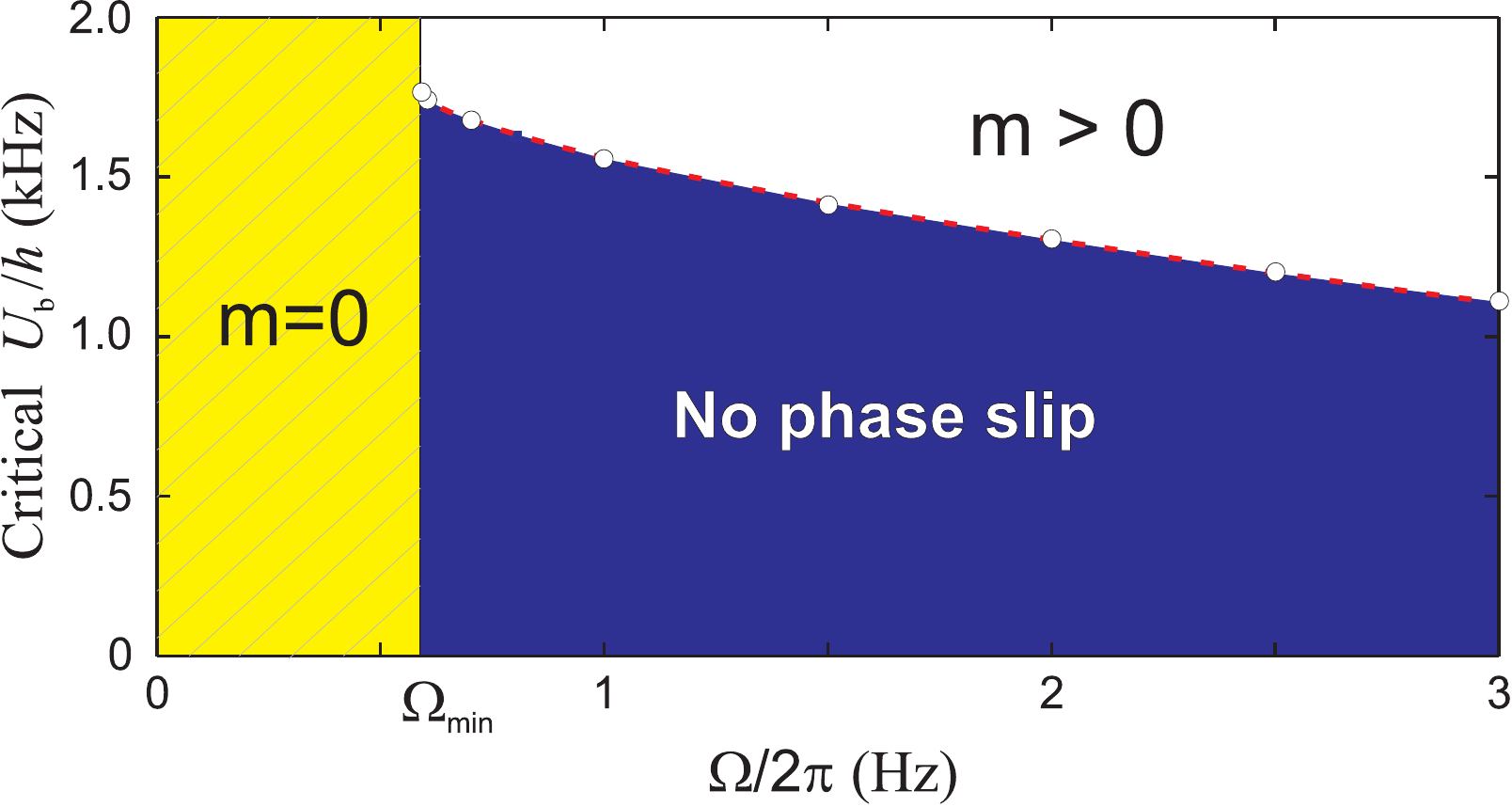}
  \caption{(Color online) Parameter regions ($\Omega$, $U_b$) where no phase slip is observed (the region shaded with blue color).  For frequencies $\Omega<\Omega_\textrm{min}\approx\frac12\Omega_0$ (shaded with gray color and crossed by the black lines region) no phase slip is possible for arbitrary barrier intensity since formation of the persistent current is energetically unfavorable. The  phase slips  $0\to 1$ are observed for $\Omega$ and $U_b$ from unshaded region. Dashed red curve is the threshold for generation of the persistent current fitted by the half-cubic parabola given by
   Eq. (\ref{halfcubic}). Open circles are the results obtained by numerical simulations of the 3D dissipative GPE. \emph{Source}: Adopted figure from paper by author \cite{yakimenko2014vortices}.}
  \label{Threshold}
  \end{figure}

We have performed in Ref. \cite{yakimenko2014vortices} a series of numerical simulations for various combinations of $U_b$ and $\Omega$.
To summarize our finding, we present  in Fig.~\ref{Threshold} the threshold for $0\to 1$ phase slip in the plane of the weak link parameters: barrier height $U_b/h$ and $\Omega/(2\pi)$.
The critical barrier height $U_b/h$ for each $\Omega$ shown in Fig.~\ref{Threshold} was obtained with accuracy not worse than 5\,Hz. In qualitative agreement with the experiment, the threshold barrier height increases when the angular velocity decreases. Particularly striking is that below a well-defined angular velocity $\Omega_\textrm{min}/2\pi \approx  0.58\,$Hz the phase slip does not occur at any barrier height $U_b$. The value of minimal angular velocity appears to be close to predictions of Ref.~\cite{nature14}: $\Omega_\textrm{min}\approx \Omega_0/2$, where $\Omega_0=\hbar/(M R^2)= 2 \pi \times  1.13\,$Hz is the rotational quantum. As it will be shown in the Sec. \ref{sec_hysteresis}, formation of the $m=1$ persistent currents becomes energetically unfavorable for $\Omega<\Omega_\textrm{min}$.

For even higher angular velocities of the rotating barrier the persistent current generation mechanism is quite different from the slow--rotation case and governed by the excitation of external surface modes, similarly to what was observed in Ref. \cite{SmallWL_arxiv14}. Worth noticing here that highly-charged persistent currents, which are generated by the rapidly-rotating barrier can be unstable with respect to serial \textit{emission} of vortex lines. During such an emission process a vortex line spirals out from the central hole of the toroid through the bulk of the condensate without weak link.

 Our theoretical results are in qualitative agreement with the experimental findings: the threshold angular velocity decreases when the barrier height increases. However, comparison of the two experimental series for the fixed barrier height $U_b$ with Fig.~\ref{Threshold} shows that the theoretically predicted $\Omega$ is higher than the experimentally measured angular velocity for the phase slip $0\to 1$.


\section{Hysteresis in a superfluid ring}\label{sec_hysteresis}
Hysteresis is widely used in electronics and it is routinely observed in superconducting
circuits. Controlled hysteresis in atomtronic circuits may prove to be a crucial feature for the
development of practical devices, as it has proved in electronics, e.g. in memory cells, digital
noise filters and magnetometers. Very recently \cite{nature14} the hysteresis between quantized circulation
states in a ring-shaped condensate with a rotating weak link has been experimentally demonstrated. It was proved, that generation of the persistent current becomes possible at angular velocity $\Omega_1$ of the repulsive barrier, but the decay of the persistent current corresponds to lower rotation rate $\Omega_2$.
The experimental results of Ref. \cite{nature14} suggest that relevant excitations involved in the observed processes are
vortices, and indicate that gain and dissipation of the energy has an important role in the dynamics
in this open-dissipative quantum system. However, the details of the process of abrupt change of the rotation state driven by a tunable rotating weak link are not elucidated yet. Furthermore, a challenging issue is that the width of a hysteresis loop $\Omega_1-\Omega_2$, obtained in the framework
of commonly used version of the dissipative GPE, appears to be substantially overestimated in comparison with the experimental measurements. It was assumed in Ref. \cite{nature14} that an agreement with the experiment will be reached in the framework
of more sophisticated models that allow for changes in chemical potential and atom number.
To describe the hysteresis process we use here the generalized 2D dissipative GPE with time-dependent chemical potential, which accounts on a decay of the number of atoms with time.

\begin{figure}[h]
\centering
\includegraphics[width=0.8\textwidth]{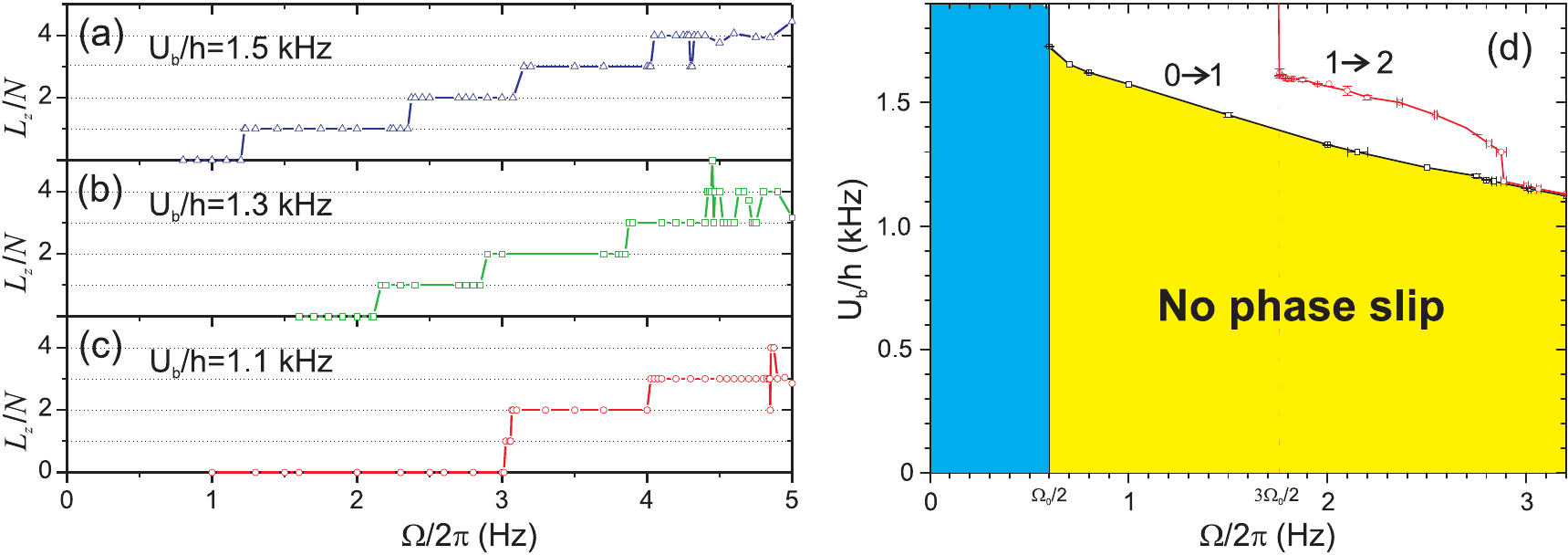}
\caption{(Color online) Results obtained in 2D model: (a)-(c)  Angular momentum per atom as a function of angular velocity for different values of the barrier height. (d) Parameter regions ($\Omega, U_b$) where no phase slip is observed (shaded with yellow color region) and the phase slips $0\to 1$ or $1\to 2$ are observed (unshaded region). The black curve indicates the edge of the region where $0\to 1$ phase slips are observed, the red curve -- edge of the region where the phase slips $1\to 2$ are observed. Formation  of the persistent current becomes energetically unfavorable for frequencies $\Omega < \frac12\Omega_0$ (region shaded with blue color).}
\label{LvsOmega2D}
\end{figure}

In the present section we use a 2D analogue of the 3D model described in Sec. \ref{sec_weaklink}.
As pointed out above, the BEC cloud is highly anisotropic in the experimental setup, thus the 2D model is supposed to properly describe the main features of the vortex dynamics.
The first task of this section is to establish the parameters of the 2D model, which gives a good quantative agreement between the details of dynamics of the vortex excitations observed in the 3D model and in the reduced 2D approximation. Having this goal in mind we use the parameters of the experiment \cite{PhysRevLett.110.025302}  and compare our findings concerning generation of the persistent current with the results of the more general 3D model, described in Sec. \ref{sec_weaklink}. We note that the problem of the dimensionality reduction in context of ring-shaped BEC has been discussed in the literature (see, e.g. \cite{PhysRevA.76.063614,PhysRevA.74.023617}), however, to the best of our knowledge, the direct comparison of the dynamics of the vortices in 3D toroidal condensate and reduced 2D annular model was not performed yet. For further chemical potential of the 2D model is related to 3D chemical potential as follows: $\mu_{3D}=\mu_{2D}+\frac12\hbar\omega_z$. Such relation is prescribed by the factorization of the $z$-coordinate (\ref{factorization}) and $\frac12\hbar\omega_z$ is nothing but the energy of the ground state in the oscillator potential. In dimensionless units the parameters of 2D model are as follows: $g_{2D}=0.016$, $N=4\cdot 10^5$, $\mu=14$. Our results obtained in the framework of (2+1)D dissipative GPE are in very good agreement with the 3D results (compare Figs. \ref{LvsOmega2D} (d)  with \ref{Threshold}). For slow rotation rate the vortices forme a dipole, which escapes from the condensate in a similar way as it happens in 3D model \cite{yakimenko2014vortices}.
At relatively rapidly rotating barrier, the dependence of the angular momentum per atom becomes irregular (see Fig. \ref{LvsOmega2D}) and it becomes difficult to find a threshold for a phase slip $m\to m+1$ with $m>1$. As is seen from Fig. \ref{LvsOmega2D} (c), (d), for higher rotation rate the thresholds for the phase slips $0\to 1$ and $1\to 2$ merge.
It is noteworthy that the higher order phase slips $m\to m+1$ for  $0\le m\le 3$ occur in our numerical simulation only above a well defined angular velocity of the weak link: $\Omega>(m+\frac12)\Omega_0$.

\begin{figure}[ht]
\centering
\includegraphics[width=0.75\textwidth]{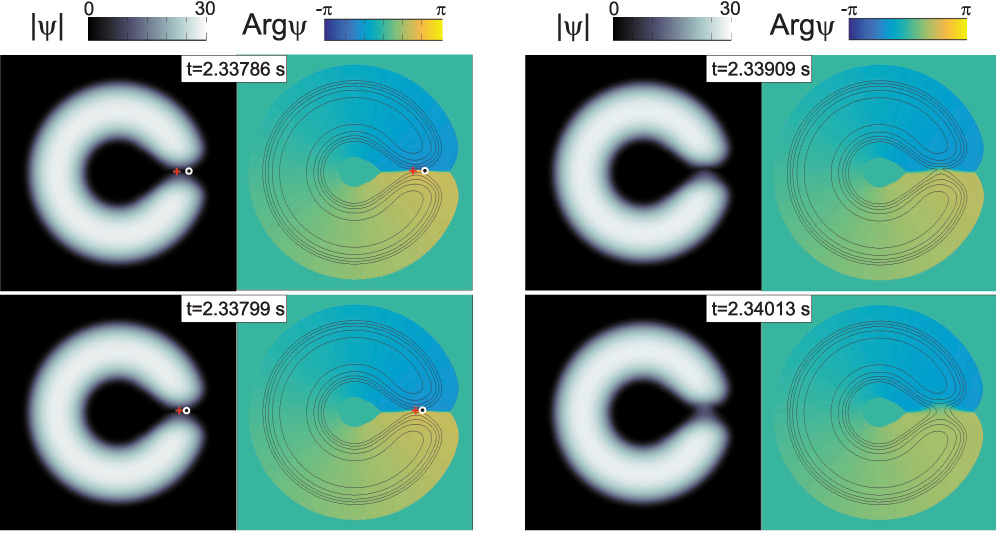}
\caption{(Color online) Decay of the persistent current driven by the non-rotating ($\Omega=0$) barrier $U_b/h=1.38$ kHz. The persistent current was prepared by the stirring procedure duration of 1.5 s. The superflow decays via annihilation of the vortex-antivortex pair in the weak link. The red cross (white circle) indicates position of the vortex (antivortex) core. Shown are $|\psi(x,y)|$ and Arg$\psi(x,y)$ for different moments of time. The size of each image is $40\times 40$ in units of $l_r$.}
\label{DecayNonRotating}
\end{figure}
\begin{figure}[ht]
\centering
\includegraphics[width=0.75\textwidth]{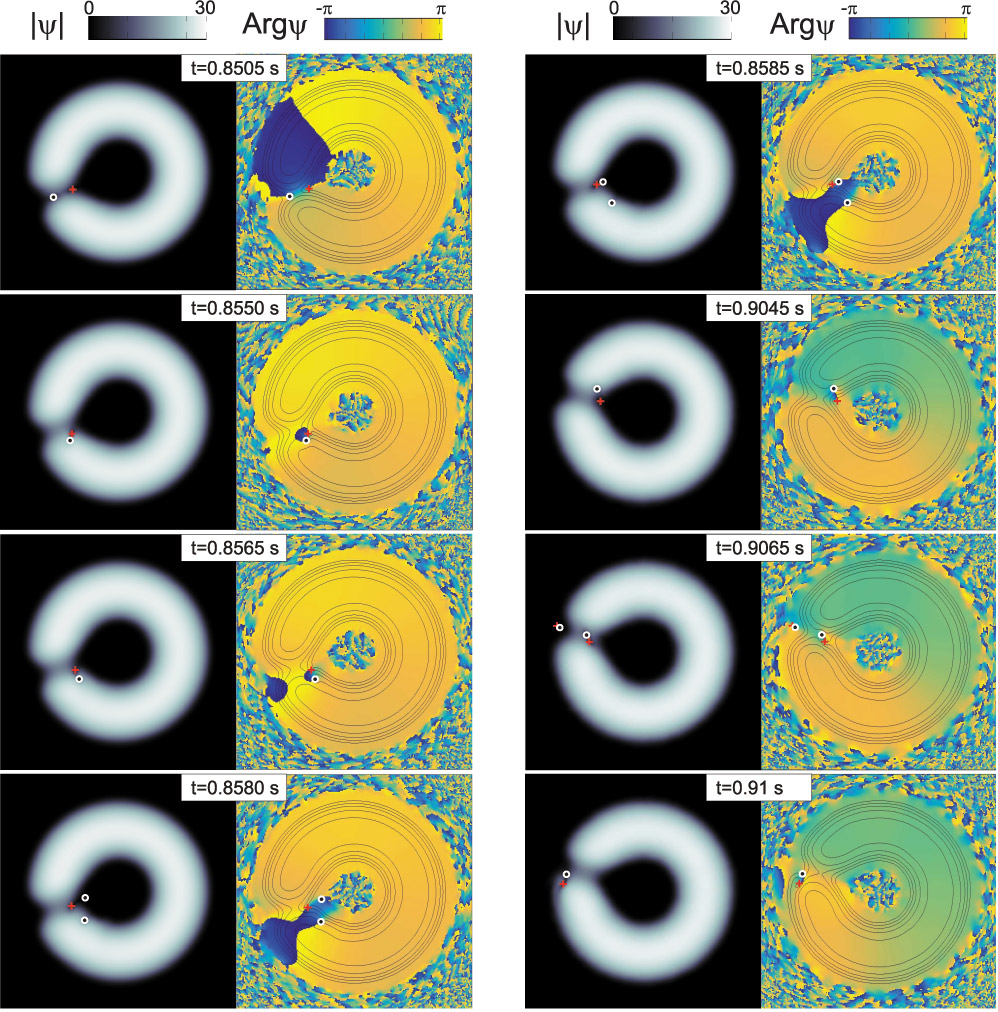}
\caption{(Color online) Decay of the persistent current driven by the rotating barrier with $\Omega_2/2\pi=-1.7$ Hz, $U_b/h=1.1$ kHz. The antivortex from the external surface and vortex from the central hole form a moving dipole. The vortex dipole penetrates into the weak link and reaches the central hole of the toroidal condensate. After series of the dissociation and annihilation with the antivortices in the central hole the vortex forms a bound pair with an antivortex and fly off through the weak link. As the result, the superflow undergoes a $1\to 0$ phase slip.  The red cross (white circle) indicates position of the vortex (antivortex) core. Shown are $|\psi(x,y)|$ and Arg$\psi(x,y)$ for different moments of time. The size of each image is $40\times 40$ in units of $l_r$.}
\label{DecayRotating}
\end{figure}

Having established a quantitative agreement between 3D and 2D models for the generation of the persistent current we considered the inverse process, a \emph{decay} of the superflow in the ring-shaped BEC by the rotating weak link.
Further we restrict our consideration to the first phase slip $0\to 1$, when the single-charged persistent current is generated, and inverse process of persistent current decay, corresponding to the phase slip $1\to 0$. Also we concentrate on the case of slow rotation rate when the barrier height is just above a threshold value for vortex generation (see Fig. \ref{LvsOmega2D} (d) and Fig. \ref{Threshold}).
 We use a state with $m=1$, which is formed by the 1.5~s stirring process  with parameters $U_b, \Omega_1$, as an initial condition for a series of simulations of another 1.5~s stirring process with the same barrier height $U_b$ and different values of angular velocity $\Omega_2$. Some typical examples of the dynamics of the persistent current decay are given in Figs. \ref{DecayNonRotating}, \ref{DecayRotating}

Let us discuss the microscopic mechanism of the persistent current generation and decay in more details.
It was pointed out in Ref. \cite{nature14}, that both generation and decay of the persistent current in a stirred ring-shaped condensate are associated with dynamics of the excitations in the form of the vortex dipoles, which is a generic feature of a superfluidity in a quasi-two-dimensional geometry. We find out that the phase slips observed in our numerical simulations are generally accompanied by formation of a \emph{moving}  vortex dipole.  As was illustrated in the Sec. \ref{sec_weaklink}  (see Fig. \ref{StremLines}) for the case of the $0\to 1$ phase slip the azimuthal velocity of the superflow is positive (forward flow) far from the weak link, but $v_\varphi<0$ inside the weak link (backward flow). The rotating weak link induces the backward component of the superflow also during the inverse phase slip $1\to 0$. When the vortex and antivortex form a bounded pair inside the rotating weak,  the linear momentum of the backward flow can be transferred into the momentum of the dipole. The attraction between  the oppositely charged vortex lines can be balanced by the Magnus force, which is induced by the superflow of the  moving vortex dipole. As the result the vortex dipole moves with linear velocity $v_d\sim \frac{\hbar}{M D}\ln(D/\xi)$, where $D$ is the dipole size (distance between the cores of vortex and antivortex), $\xi$ is the healing length.

As it is known (see, e.g. \cite{2014JLTP..175..189T}), the vorticity energy $E_v(D)$ of the vortex dipole in an inhomogeneous condensate decays both for large dipoles (when $D$ is of order of the condensate size so that the vortex cores appear at the low-density region) and for small dipole sizes (when $D$ is of order of the vortex core radius $\xi$). Therefore, the contribution to the condensate energy $E_v(D)$ of a vortex-antivortex pair
 as a function of dipole size $D$
 must have a maximum, which defines the dissociation size $D_m$ of the dipole: $d E_v/d D|_{D=D_m}=0$. Formation of the vortex dipole can be described as follows:
When the barrier height grows, the width of the condensate in the weak link decreases so the distance $D$ between vortex and antivortex from the internal and external side of the annulus decreases. For a compact vortex-antivortex pair with $D<D_m$ formation of the bound state in the form of vortex dipole becomes energetically preferable.

As was pointed out above, the moving dipole can be supported against 2D reconnection (collapse) by the Magnus force. However, if the rotation rate of the weak link is very low, the vortex dipole, emerging during phase, slip is too slow to form a stable bound state. In this case the collapse inside the weak link is observed (see Fig. \ref{DecayNonRotating}), which is consistent with previous theoretical findings  \cite{PhysRevA.80.021601,Piazza2013} concerning persistent current decay in a toroidal condensate with a non-rotating tunable weak link.

Therefore during the phase slip the linear momentum of the backward flow, induced by rotating weak link, is picked up by the moving dipole.  Further evolution of the moving dipole can be accompanied by different processes: by dipole dissociation; by topological charge exchange of the moving dipole with the "ghost vortices" at the periphery of the condensate; by reversible transformation of the dipole into a \emph{gray soliton} \cite{2013JPhB...46l5302P,2014arXiv1403.4658K} or by annihilation of the dipole with sound pulse emission. Illustration of some of these processes is given in Fig. \ref{DecayRotating}, however the detailed description of the complex dynamics of the vortices in the stirred toroidal condensate
 is beyond the scope of the present work.

The remarkable feature associated with generation and decay of the persistent current is the
hysteresis effect \cite{nature14}.
Hysteresis is a general property of systems where the energy has multiple local minima separated by an energy barrier. For a Bose--Einstein condensate (BEC) in a ring-shaped trap, these minima represent
stable flow states of the system, and their energies depend on the
applied rotation rate of the trap $\Omega$ and the amplitude $U_b$ of a rotating repulsive perturbation (weak link), which induces the rotation.
\begin{figure}[h]
\centering
\includegraphics[width=0.75\textwidth]{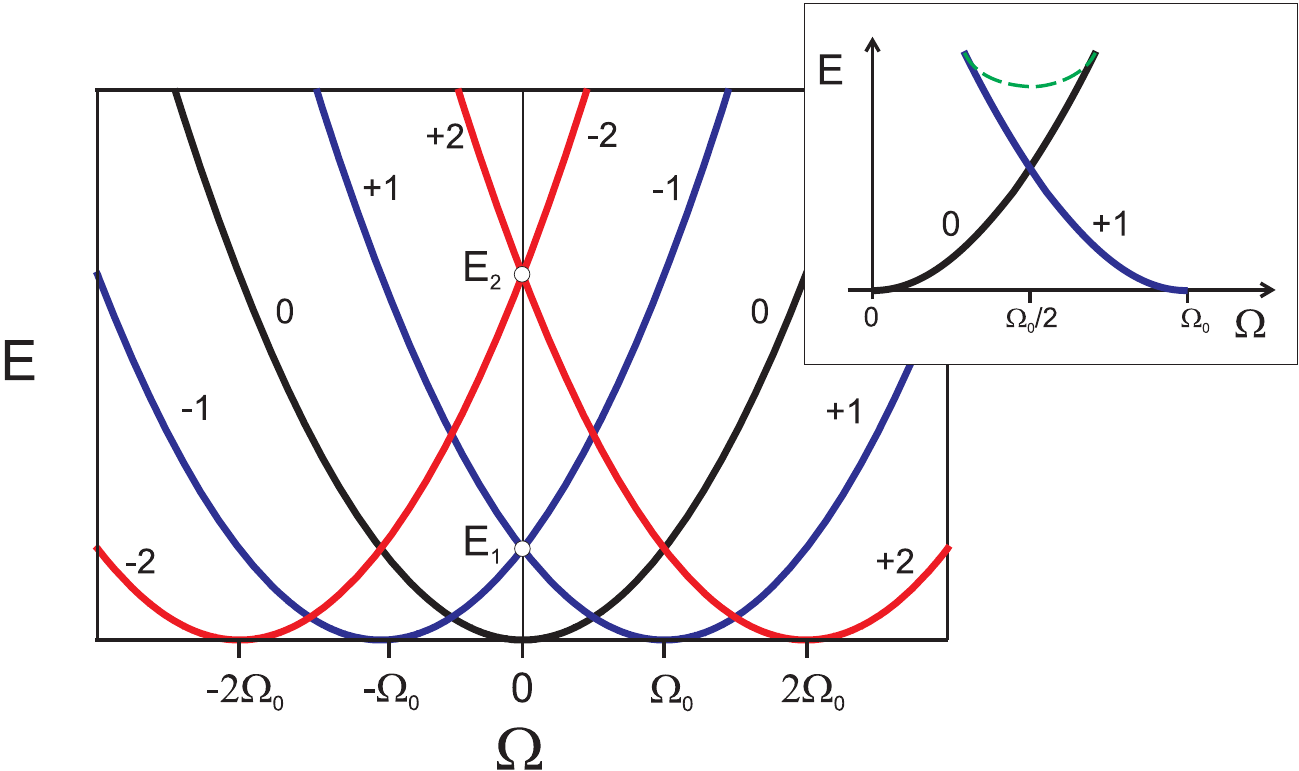}
\caption{(Color online) Single-particle energy levels in the rotating frame as a function of $\Omega$ for winding numbers $m$ indicated near the curves. Inset shows the energy of $m=0$ (solid black line) and $m=1$ states (solid blue line). The dashed green line depicts the energy of the barrier, which separates these metastable states.}
\label{spectrum}
\end{figure}

As known (see, e.g. \cite{PhysRevA.87.013619,PhysRevA.66.063603})  the noninteracting single-particle energy levels, shown in Fig. \ref{spectrum}, are periodic functions of $\Omega$ with period $\Omega_0$.
The single-particle spectrum in a 1D ring of radius $R$ can be described in the rotating frame as follows: $E=\frac{1}{2}\hbar\Omega_0\left(m-\Omega/\Omega_0\right)^2,$
where $\Omega_0=\hbar/(MR^2)$ is the characteristic scale of rotation in
the system.
With the addition of interactions, an energy barrier appears, separating $m$-charged metastable flow states (see the inset in Fig. \ref{spectrum}).  This barrier stabilizes the flow, making the persistent current very long-lived.

\begin{figure}[ht]
\centering
\includegraphics[width=0.55\textwidth]{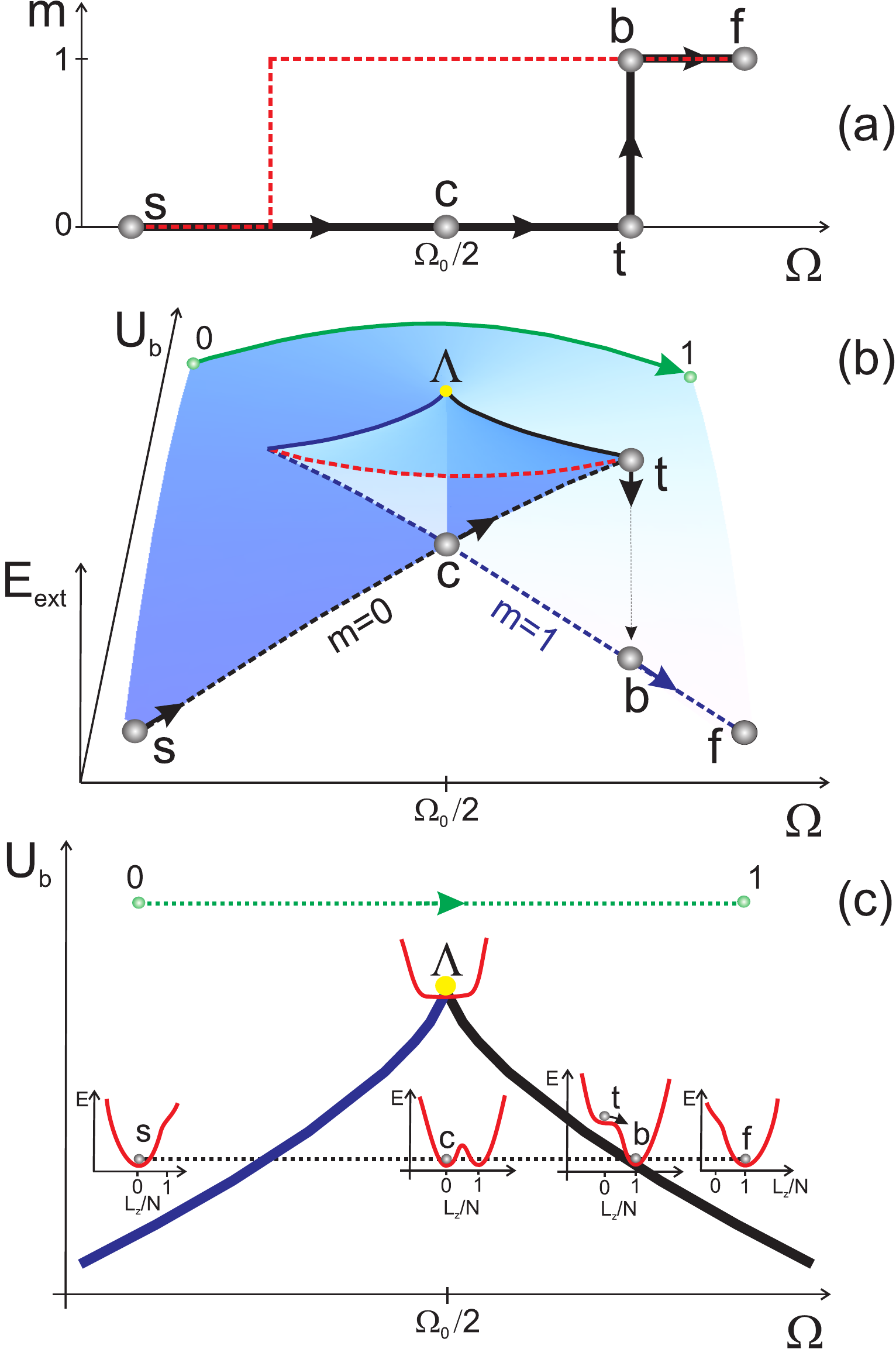}
\caption{(Color online) (a) Schematic of the angular momentum per atom $m=L_z/N$ as a function of angular velocity $\Omega$ at fixed barrier height $U_b$. The solid black  line (for the  phase slip  $0\to 1$) and dashed red line (for the inverse transition $1\to 0$) form the hysteresis loop. (b) Three-dimensional depiction of the a swallow-tail structure of the energy extreme as a function of control parameters $\Omega$ and $U_b$. The slice of this surface at fixed $U_b$ presents the energies of the three extreme:  two minima (black dashed line for $m=0$ and blue dashed line for $m=1$), and one maximum (red dashed line).  (c) The schematics of the energy landscape as a function of $m$ for different $\Omega$. The catastrophe set: inside the cusp there are three extrema: two minima for $m=0$ and  $m=1$, which are separated by a maximum. For the control parameters outside the cusp there is only one minimum and no catastrophe appears at the trajectory indicated by dashed green line.}
\label{catastrophe}
\end{figure}
\begin{figure}[ht]
\centering
\includegraphics[width=0.7\textwidth]{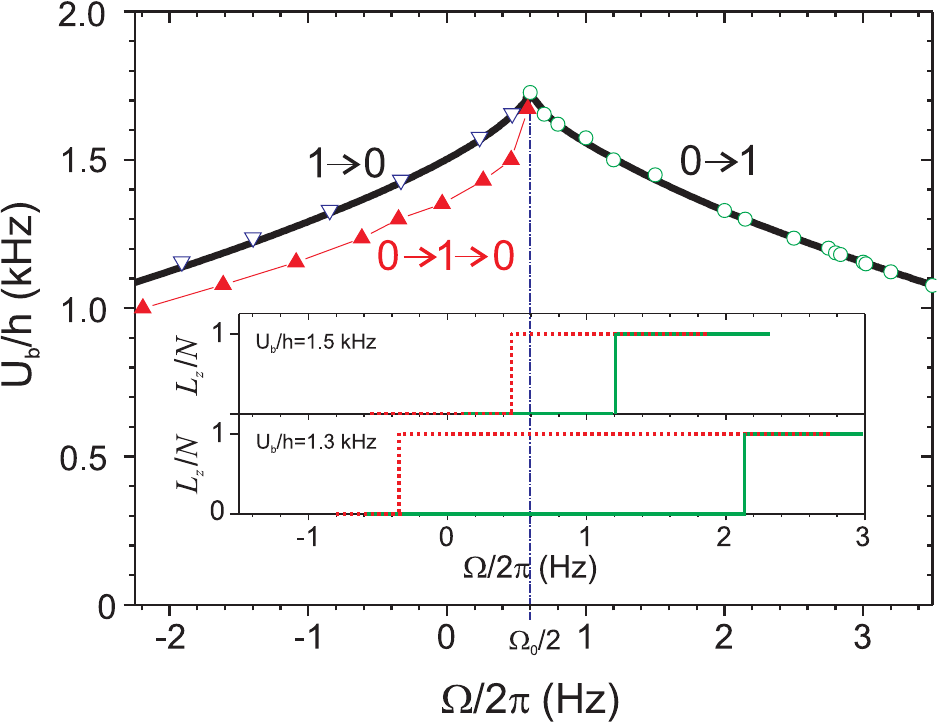}
\caption{(Color online) The critical lines of the cusp catastrophe fitted by the half-cubic parabola [see Eq. (\ref{halfcubic})] are shown by solid black lines. Open green circles indicate  the numerical results for the boundary of the region with $0\to 1$ phase slips. Open blue triangles indicate  the numerical results for the boundary of the region with $1\to 0$ phase slips. Duration of the stirring procedure is 1.5 s for the both series, the initial number of atoms is the same. The red filled triangles indicates the boundary for the backward phase slip $1\to 0$; the total duration of the two-stage procedure $0\to 1\to 0$ is 3 s.   Inset presents two examples of the hysteresis loops for $U_b/h=1.3$ kHz and $U_b/h=1.5$ kHz. Note that due to considerable decay of the number of atoms during the simulation time the hysteresis loops shrink and become substantially asymmetric with respect to the frequency $\frac12\Omega_0$ indicated by blue dashed-dotted line.}
\label{hysteresis}
\end{figure}
To gain a better insight into the properties of the generation and decay of the persistent current it is useful to consider an energy landscape of a rotating superfluid ring.
The energy of the condensate as a function of the angular momentum per particle $m=L_z/N$ are sketched in Figs.~\ref{catastrophe} (c) for different values of $\Omega$. It turns out that for some range of control parameters $\Omega$ and $U_b$ the energy landscape exhibits  two local minima at integer $L_z/N$ (here we consider $m=0$ and $m=1$ states). The values of the energy at minima for different values of $\Omega$ and $U_b$ form a self-intersecting surface, which is shown schematically in  Fig.~\ref{catastrophe} (b). Obviously, the presence of two local minima at a given value of control parameter assures that there is a maximum separating them.  The barrier state (energy of the maximum) is also shown in Fig.~\ref{catastrophe} (b) forming the top of the ``swallow tail''.  The slice of the surface $E_\textrm{ext}$ at fixed $U_b$ in Fig. \ref{catastrophe} (b) represents the energy of the maximum  by dash-dotted red line, and the energy at $m=0$ ($m=1$) minimum by dashed black (blue) line (see also the inset in Fig.~\ref{spectrum}).  At the point labeled by $t$ in Fig.~\ref{catastrophe}, one of the local minima  (with $m=0$) meets up with the maximum, and they both disappear. The points where the maximum of the energy merges with the minimum at $m=0$ ($m=1$) form the solid black (blue) line.
The projections of these lines onto the plane $(\Omega, U_b)$ are shown in Fig.~\ref{catastrophe} (c) and Fig.~\ref{hysteresis}.
The region with three extrema of the energy landscape is bounded from above by these curves.
The ``swallow-tail'' energy spectrum is associated with \emph{a cusp catastrophe} with a typical for such catastrophe half-cubic line of the catastrophe:
\begin{equation}\label{halfcubic}
(\Omega-\Omega_\Lambda)^2=\alpha^6 (U_\Lambda-U_b)^3,
\end{equation}
 where $(\Omega_\Lambda, U_\Lambda)$ is the coordinate of the cusp point $\Lambda$ indicated in Fig. \ref{catastrophe} (b),(c) and $\alpha$ is a constant.
As was pointed out above, the energy spectrum of the condensate in the ring is a periodical function of the angular velocity with period $\Omega_0$, where $\Omega_0$ is the quantum of rotation. From the symmetry of the energy spectrum it follows, that the energy spectrum branches for $m=0$ and $m=1$ cross at $\Omega=\frac12\Omega_0$. Thus, the cusp point in catastrophe set corresponds to $\Omega_\Lambda=\frac12\Omega_0$.  It is remarkable that the function (\ref{halfcubic}) fits the threshold for $0\to 1$ transition surprisingly well (see Fig. \ref{hysteresis} and Fig. \ref{Threshold}).

A ``swallow-tail'' structure of the energetic spectrum results in a hysteretic response of nonlinear system to  variations of a control parameter. The transitions $0\to 1$ and
$1\to 0$ occur at different values of $\Omega$ and form a hysteresis loops illustrated in Fig. \ref{catastrophe} (a).  The hysteresis exists for that region of $(U_b,\Omega)$, where there are three extrema of the energy landscape. It is instructive to consider the response of the system on the variation of the angular velocity of the weak link at fixed barrier amplitude. Let us describe an example of the generation of the persistent current. We start at a ground state (point $s$ in Fig. \ref{catastrophe}), when there is only one energy minimum at $m=0$. When $\Omega$ grows we cross the blue cusp curve and the second local minimum at $m=1$ appears. The energy of two local minima become equal at $\Omega=\Omega_0/2$ (point $c$). At the point labeled by $t$ in Fig. \ref{catastrophe}, the local minima  at $m=0$ disappear and  the system jumps to the state $b$ with a minimum at $m=1$. This abrupt transition corresponds to the phase slip $0\to 1$ shown in Fig. \ref{catastrophe} (a). The local maximum of the energy corresponds to an unstable solution, which is associated with the vortex excitations.  If $\Omega$ grows even further, the local minimum at $m=1$ also disappears and phase slip $1\to 2$ is observed.


The Figure \ref{hysteresis} summarizes our findings concerning the boundaries of the regions with the phase slip: open green circles give the boundary for the $0\to 1$ phase slips [shown also in Fig. \ref{LvsOmega2D} (d)], filled red  triangles represent the edge of the region with $1\to 0$ phase slips. The filled red triangles correspond to the two-stage stirring procedure $0\to 1\to 0$: the $m=1$ persistent current is generated during 1.5 s of the first stirring stage, the persistent current decay via $1\to 0$ phase slip during next 1.5 s.

Consideration of the energy landscape of the conservative system suggests the symmetry of the hysteresis loop with respect to the frequency $\Omega_0/2$ [see Fig. \ref{catastrophe} (a)]. In a sharp contrast to these expectations a considerable asymmetry of the hysteresis loop, obtained by direct numerical simulations of the dissipative dynamics of the condensate, is clearly seen in the numerically obtained hysteresis loops (see the inset in Fig. \ref{hysteresis}). Note that the $1/e$  lifetime of the condensate is specified to be 10 s, that results in a decay of about a quarter of the initial number of atoms during the 3 seconds of simulation time. Thus, the asymmetry of the hysteresis loop in this case can be explained by a considerable asymmetry in the initial values of the number of atoms for the forward ($0\to 1$) and backward ($1\to 0$) branches of the hysteresis loop.

In support of this conjecture we have verified, that the hysteresis loop becomes symmetric providing the same duration of the stirring procedure and equal values of the total number of atoms in the initial state for generation of persistent current via the $0\to 1$ phase slip and for decay of the persistent current via $1\to 0$  phase slips. The open blue triangles in the Fig. \ref{hysteresis} correspond to the threshold of the $1\to 0$ phase slip, when the stationary solution of the GPE with $m=1$ was used as initial condition for the stirring procedure duration of 1.5 s.

It is remarkable, that the experimental hysteresis loops \cite{nature14} are symmetric with respect to $\Omega_0/2$.
The point is that to observe a symmetric hysteresis in the phase slips, in Ref. \cite{nature14} a two-step experimental
sequence was used. After condensing the atoms into the ring
trap, the BEC is prepared in either the $m=0$ or the $m=1$ circulation
state by either \emph{not rotating} the weak link or by rotating it.
The weak link was then rotated at various angular velocities, for an additional 2 s.

It was underlined in Ref. \cite{nature14}  that there is a large
discrepancy between the models used in \cite{nature14} and the experimental observations: the width of the hysteresis loop, obtained from the numerical simulations of the dissipative GPE with time-independent chemical potential (and consequently, a constant number of atoms), was much greater, then experimentally measured  width of the hysteresis loop.
As was pointed out above, the decay of the number of atoms leads to shrinking of the width of the hysteresis loop. Thus, it is of interest to account on this dissipative effect using time-dependent chemical potential in the framework of the stirring procedure and parameters of the experimental setup \cite{nature14}.
As it follows from our theoretical results reported in the present paper, the correspondence between theoretical predictions and experimental observations can be radically improved, by accounting for the decay of the number of atoms with time. This work is now in progress and the results will be published elsewhere.

\section{Conclusions}\label{Conclusions}
In the present paper we overview our three recent works \cite{PRA2013R,SmallWL_arxiv14,yakimenko2014vortices}, which provide a theoretical description of three different experiments \cite{Beattie13,PhysRevLett.110.025302,Wright2013} with persistent current in toroidal condensate. In addition we present our new results related to hysteresis in the ring-shaped condensate, which was very recently demonstrated experimentally \cite{nature14}.

Existence of a long-lived persistent current corresponds to stable quantum vortex pined in the central hole of the toroidal condensate. The lifetime of this vortex line in a single-component condensate ranges up to two minutes and restricted only by the lifetime of the condensate itself.
However, in the case of the two-component (spinor) BEC the stability properties of the persistent current dramatically change \cite{Beattie13}.
In Ref. \cite{PRA2013R} we have found that phase slips occur as the result of the simultaneous
action of two factors: an azimuthal symmetry-breaking instability and a dissipation effect caused by the interaction of
the condensate atoms with the thermal cloud. If the number
of atoms in the minor spin component becomes comparable
to the number of atoms in the major component, the nonlinear
repulsive intercomponent interaction leads to a separation of
the spin components through an azimuthal symmetry-breaking
instability. As a result of this instability, one can find regions in
the condensate annulus with reduced density of one component
filled by atoms of the other component. These regions serve
as ``gates'' for the vortex lines where they readily cross the
annulus, and as a consequence, the winding number of the
superflow is reduced by one unit.

We have discussed our recent theoretical results concerning generation of the quantized vortices by a rotating repulsive barrier.
It is remarkable that for low rotation rate both a weak link (a barrier which is wider than the condensate annulus) \cite{PhysRevLett.110.025302} and a small (in comparison with the width of the annulus) stirrer used in Ref. \cite{Wright2013} drive the phase slip by excitation of the vortex dipoles. However, the microscopic mechanisms of the phase slip are \emph{qualitatively different} for these two cases. [see Fig. \ref{SchematicStirring}]
As was shown in Ref.~\cite{SmallWL_arxiv14}, a small stirrer at low rotation rate excites vortex-antivortex pairs near the center of the rotating barrier in the bulk of the condensate. Then the pair undergoes a breakdown and the antivortex moves spirally to the external surface of the condensate and finally decays into elementary excitations, while the vortex becomes pinned in the central hole of the annulus adding  one unit to the topological charge of the persistent current. For the case of the wide barrier (weak link) one observes the following: A vortex line enters the weak link from the outside and travels towards the central hole. While the vortex traverses the weak link, a negatively-charged anti-vortex line from the central hole and the incoming vortex approach each other and create a vortex-antivortex dipole. This coupled pair of vortices circles clockwise inside the central hole
until it reaches the region of the weak link.  The moving dipole usually escapes from the central hole and finally decays, however, during the stirring time the vortex dipole can  several times enter and leave the central hole of the ring through the weak link. Though the total charge of the escaping vortex dipole is equal to zero, but each time an external vortex (or antivortex) enters the condensate it adds (or takes away) one unit of topological charge to the global persistent current.

Using  the 2D dissipative GPE we have investigated a hysteresis effect of the persistent current in a toroidal BEC, driven by a rotating weak link.  We have revealed that both generation and decay of the persistent current are  driven by dynamics of the moving vortex dipoles.
A vortex dipole emerging during the phase slip exhibits a complicated evolution, which can be accompanied by the dipole dissociation,
by reversible transformation of the dipole into a gray soliton or by annihilation of the dipole with sound pulse emission. The moving dipole finally decays into elementary excitations at the boundary of atomic cloud. Our approach accounts for the decay in the total number of atoms and leads to considerable shrinking of the hysteresis loop for generation and decay of the persistent current. We believe that our results can be useful for theoretical explanation of the ongoing experiments with 'atomtronic' circuits.

\section{Acknowledgment}
A.Y. acknowledges the hospitality of the Physikalisch-Technische Bundesanstalt
(PTB), where this work was commenced. S.V. is grateful to the Swiss National Science Foundation
(individual grant N IZKOZ2\verb=_=154984) and to  Prof. Ruth Durrer for her support and kind hospitality.

\newpage

\bibliography{bibliography2}
\end{document}